\documentclass[twocolumn,superscriptaddress,longbibliography,aps,pra,floatfix]{revtex4-2}

\usepackage{graphicx}
\usepackage{bm}
\usepackage{color}
\usepackage{amsmath}
\usepackage{amssymb}
\usepackage{epstopdf}
\usepackage{gensymb}
\usepackage{enumitem}
\usepackage{float}
\usepackage[urlcolor=blue,colorlinks=true,citecolor=blue,linkcolor=blue,pdfstartview={FitH},bookmarks=false]{hyperref}
\usepackage{xfrac}
\usepackage{amsmath,bm}

\begin{document}

\title{Charge ordering mechanism in silver fluoride AgF$_2$} 

\author{Mariana Derzsi} 
\affiliation{Advanced Technologies Research Institute, Faculty of Materials Science and Technology in Trnava, Slovak University of Technology in Bratislava, J. Bottu 25, 917 24,Bratislava, Trnava, Slovakia}
\affiliation{Center of New Technologies, University of Warsaw, Zwirki i Wigury 93, 02089 Warsaw, Poland}

\author{Kamil Tok\'{a}r} 
\affiliation{Advanced Technologies Research Institute, Faculty of Materials Science and Technology in Trnava, Slovak University of Technology in Bratislava, J. Bottu 25, 917 24,Bratislava, Trnava, Slovakia}
\affiliation{Institute of Physics, Slovak Academy of Sciences, D\'{u}bravsk\'{a} cesta 9, 845 11 Bratislava, Slovakia}

\author{Przemys\l{}aw Piekarz} 
\affiliation{Institute of Nuclear Physics, Polish Academy of Sciences, Radzikowskiego 152, 31342 Krak\'{o}w, Poland}

\author{Wojciech Grochala} 
\affiliation{Center of New Technologies, University of Warsaw, Zwirki i Wigury 93, 02089 Warsaw, Poland}

\begin{abstract}

Using Density Functional Theory, competition between Mott-Hubbard and intervalence charge transfer mechanism of electron localization is revealed in AgF$_2$, an important silver analogue of oxocuprates. 
We show that at reduced temperature and electron correlations AgF$_2$ becomes metallic and dynamically unstable in respect to CDW phonon modes that promote charge ordering. The CDW instability is closely related to Kohn anomaly and Fermi surface nesting.
The long advocated KBrF$_4$ type CDW Ag$^{1+/3+}$F$_2$ structure and its facile transformation to the ground state Ag$^{2+}$F$_2$ phase is explained. Our results point out to intimate interplay between lattice, charge and spin degrees of freedom in seemingly simplistic binary metal fluoride.
\end{abstract}

\maketitle
 
Copper oxide superconductors exhibit a number of competing phases when the antiferromagnetic (AFM) Mott insulating state is doped by charges~\cite{Fradkin2015,Keimer2015}. Since the first observation of the striped phase in the CuO$_2$ planes~\cite{Tranquada1995},
growing evidence of charge order (CO) existing in all families of cuprates has been accumulated~\cite{Chang2012,Comin2014,Tabis2014,Neto2015}.  
In most cases, the low-temperature CO phases are observed at moderate doping levels and they are characterized by a 2D charge density wave (CDW) 
with a modulation vector $q_\text{CDW}$~\cite{Frano2020}. The CDW is stabilized by ionic displacements and often accompanied by a phonon anomaly observed 
at the wave vector $q_\text{CDW}$~\cite{LeTacon2014,Miao2018}, thus showing similar behavior as the prototypical CDW systems~\cite{Weber2011,Zhang2014}.
Despite of extensive studies, understanding of CO formation and its role in high-$T_c$ superconductivity
is still far from complete due to complexity of materials, doping dependence, chemical disorder, and many-body effects~\cite{Frano2020,Zhang2020}.  

It was demonstrated that AgF$_2$, also known as $\alpha$-AgF$_2$ ($Pbca$), is an excellent silver analogue of the parent compound of high-$T_c$ superconducting cuprates~\cite{Grochala2001}. 
It consist of layers of spin-1/2 ions with formally Ag($d^9$) electronic configuration -- unpaired electrons reside in 
the $d_{x^2-y^2}$ orbitals and couple antiferromagnetically via superexchange mechanism that involves F($p$) orbitals~\cite{Grochala2001,Gawraczynski2019}. 
In contrast to flat CuO$_2$ layers, the neutral layers forming a square lattice in AgF$_2$ are severely buckled~\cite{Fisher1971,Grzelak2017,Sanchez2021}. 
Elimination of the buckling could result in an AFM coupling that surpasses the cuprates~\cite{Gawraczynski2019}. 
Assuming a magnetically driven mechanism, such large coupling could potentially lead to superconducting critical temperatures higher than those exhibited by cuprates
~\cite{Grzelak2020}.

The most recent experimental and theoretical study on the interband excitations in AgF$_2$ demonstrates that  AgF$_2$ is at the verge of a charge transfer instability~\cite{Bachar2021}, while independent experimental studies reported observation of a metastable 
diamagnetic $\beta$ phase interpreted as  charge-ordered Ag$^{1+}$Ag$^{3+}$F$_4$ that fast-transforms to the $\alpha$-Ag$^{2+}$F$_2$ structure~\cite{Shen1999}. 
It was proposed that $\beta$-AgF$_2$ takes on the KBrF$_4$ type structure with the $I4/mcm$ symmetry~\cite{Muller1987,Shen1999,Graudejus2000} and indeed the recent theoretical study demonstrated dynamical stability
of AgF$_2$ in the KBrF$_4$ lattice~\cite{Tokar2021}. It was also explained that the orthorhombic $\alpha$ and the KBrF$_4$ type $\beta$ phases both emerge from fluorite type structure via distinct order parameter, $i.e.$ freezing out a different phonon mode~\cite{Tokar2021}.
The Ag$^{2+}$F$_2$ layers formed by edge-sharing of square-planar [Ag$^{2+}$F$_4$] units in the $\alpha$ phase are disintegrated and replaced by
layers of isolated square planar  [Ag$^{3+}$F$_4$] units and Ag$^{1+}$ cations alternating along the tetragonal $c$ direction in the KBrF$_4$ type lattice. 
Despite the above described differences, both crystal structures are intimately related and differ only in displacements 
of the fluorine atoms, while the volume of the KBrF$_4$ type unit cell is doubled relative to the $\alpha$ phase. 
This is reminiscent of the situation in the oxocuprates, where the charge and spin ordered domains are related by displacements of oxygen atoms in the CuO$_2$ layers~\cite{Zhang2020}.

The existence of the $\beta$ phase and its transformation to $\alpha$ clearly points to a competing AFM and CDW orders
with different mechanisms of electron localization in AgF$_2$. 
Furthermore, it suggests that the CO phase is metastable relative to the spin-ordered AFM phase and quickly destroyed thermally via
the inter-valence charge-transfer (IVCT), Ag$^{1+}+$ Ag$^{3+}\rightarrow$ 2Ag$^{2+}$.
Most importantly, it indicates that CO AgF$_2$ phase may be stabilized already in an undoped regime in contrast to the superconducting copper oxides, 
where the CO appears at moderate doping levels.

The main goal of this Letter is to investigate the mechanism of charge ordering in AgF$_2$ and its possible relation to Kohn anomaly and Fermi surface nesting. 
We show that at reduced temperature and electron correlations $\alpha$-AgF$_2$ becomes metallic and dynamically unstable in respect to CDW phonon modes that promote CO. 
We also show that the long advocated KBrF$_4$ type structure of the observed charge-ordered $\beta$-AgF$_2$ naturally emerges from the $\alpha$ phase due to a phonon-induced CDW state when exact exchange is accounted for within the hybrid DFT. 
Our study shows that transition from the AFM to CO state via the metallic state in AgF$_2$ is in line with the Peierls’ picture (extended to higher dimensions) based on the concept of Fermi surface nesting. It also reveals importance of electron correlations for stabilization of both, the spin and the charge ordered AgF$_2$ phases.

The proper insulating state with the AFM order can be obtained in $\alpha$-AgF$_2$ only if local electron interactions are included beyond the standard DFT approximation,
within such methods as hybrid functionals or DFT+U approach~\cite{Kurzydlowski2018,Gawraczynski2019,Tokar2021}.
The electronic band structure calculated with hybrid DFT shows the insulating gap of 2.3 eV~\cite{Gawraczynski2019}. 
A similar electronic state was obtained within DFT+U method with the realistic value of parameter $U_\text{Ag}=5$ eV,
however, the band gap is about two times smaller (1.17~eV)~\cite{Tokar2021}. Calculations of phonon dispersions performed by DFT+U method show dynamical stability of $\alpha$-AgF$_2$ with the AFM ground state~\cite{Tokar2021,Grzelak2017}. 

\begin{figure}[t!]
\includegraphics[scale=0.17]{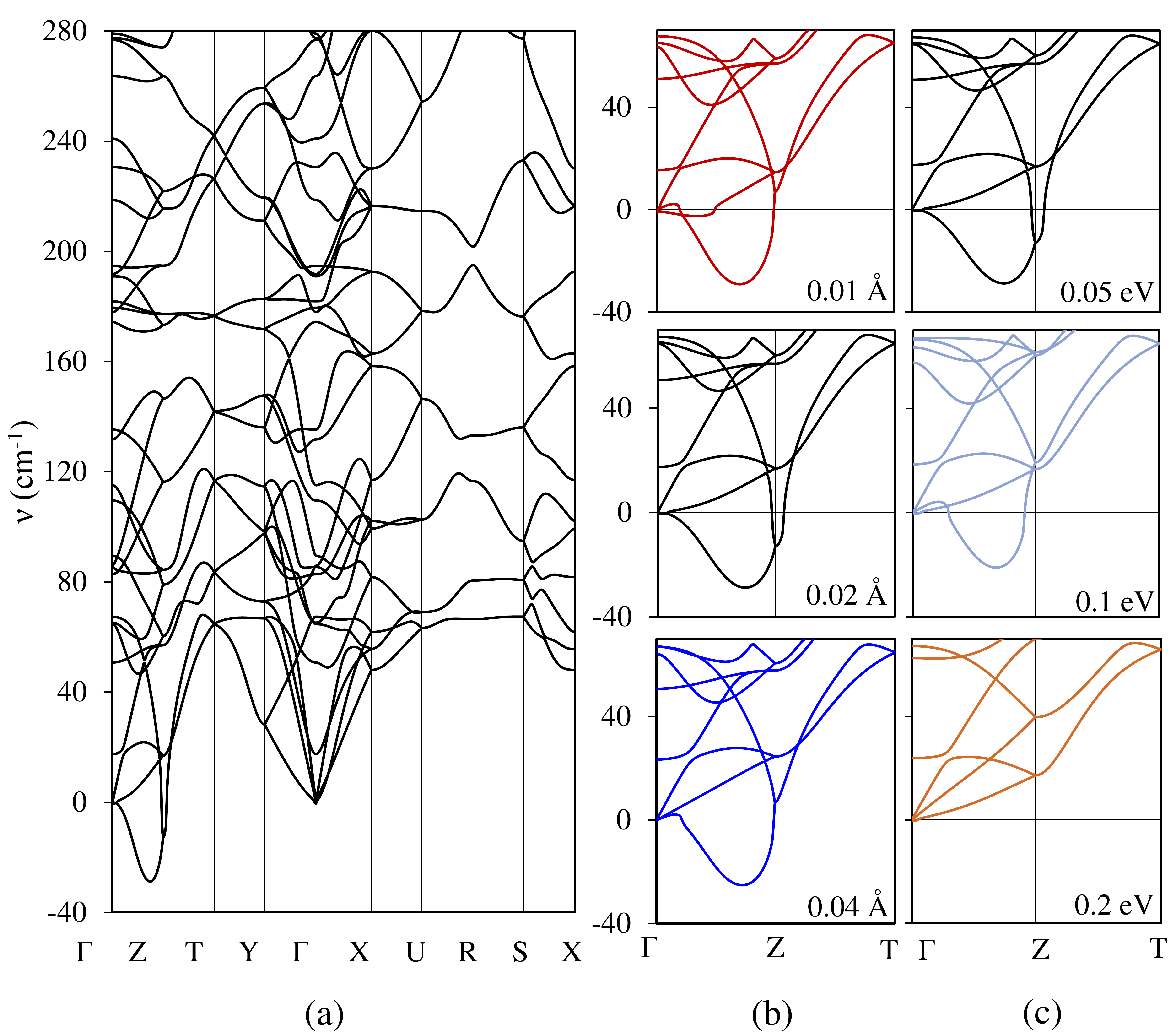}
\caption{Phonon dispersion relations in AgF$_2$ (a), and impact of the atomic displacement (b) and electron smearing width (c) on the C-CDW and I-CDW modes along the $\Gamma$-Z direction calculated for the non-magnetic state in the DFT limit. The value of the displacement and the smearing width is indicated in the  lower right corner of each plot in (b) and (c), respectively.}
\label{fig1}
\end{figure}

\begin{figure}[t!]
\includegraphics[scale=0.25]{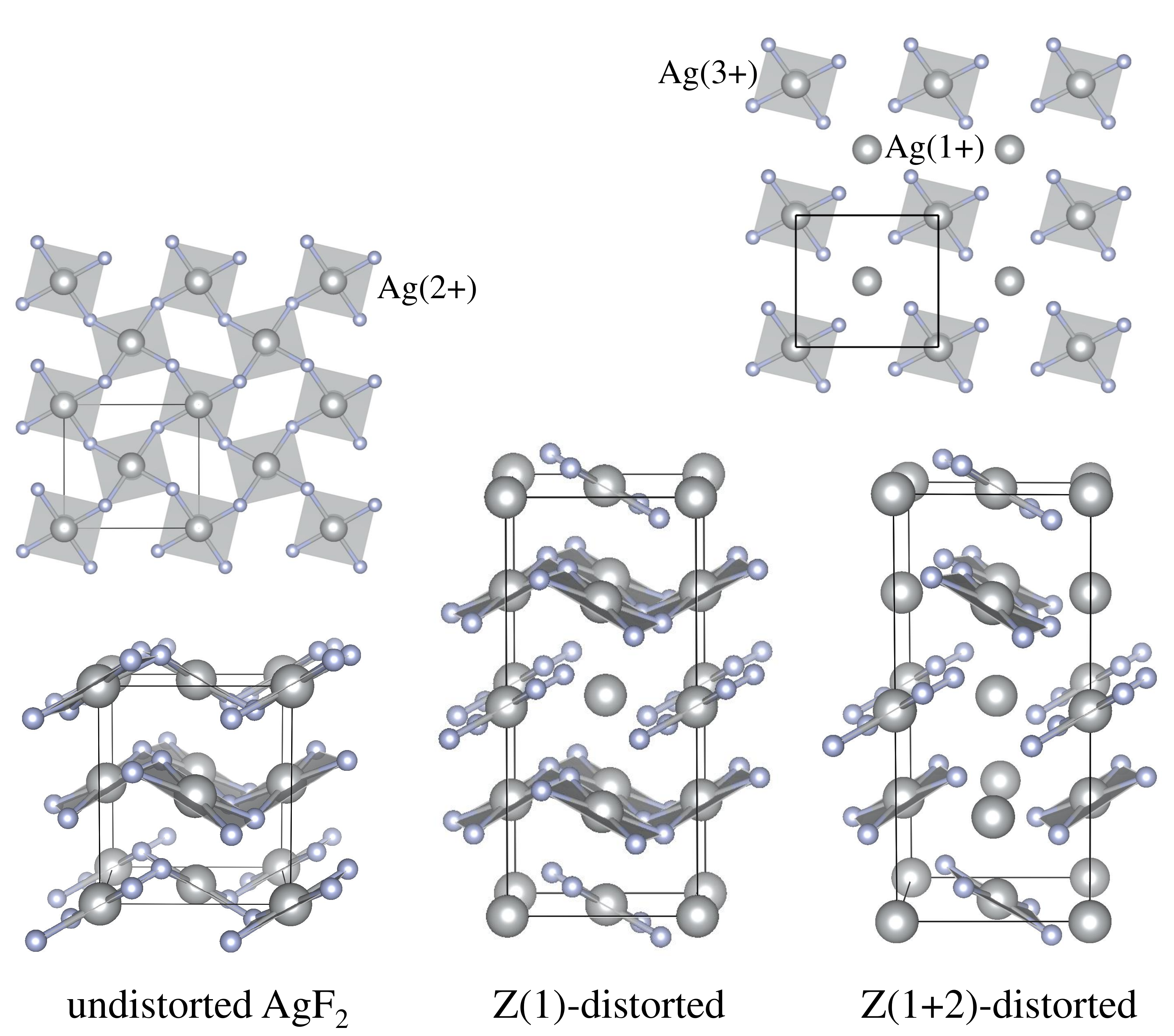}
\caption{Charge disproportionation and cell doubling induced by the doubly degenerate phonon mode at the Z point: undistorted AgF$_2$ (left), distorted by one Z(1) arm (AFM-CDW phase, middle) and both Z(1+2) arms of the Z mode (CDW phase, right). Large balls - Ag atoms, small balls - F atoms.}
\label{fig2}
\end{figure}

In the DFT regime (U=0) the electronic state is metallic without any magnetic order and
a phonon instability in a form of a sharp dip reminiscent of a Kohn anomaly develops at the Z point 
as it is shown in Fig.~\ref{fig1}(a). This is a doubly degenerate mode with the dominant contribution of F atoms. 
Both phonon branches at the Z point cause bond disproportionation within the AgF$_2$ layers due to the [AgF$_4$] breathing motion 
accompanied by doubling of the unit cell along the direction perpendicular to the layers (Fig.~\ref{fig2}). 
While all four Ag-F bonds elongate for one silver atom, they contract for nearest four silver atoms that share the same fluorine atoms. 
Consequently, all four Ag-F bonds shorten for half of the Ag atoms and elongate for the other half within the layer. 
At positions with shortened bonds, the valence corresponds to Ag$^{3+}$, while at those with elongated bonds to Ag$^{1+}$.
Thus, this breathing mode mediates CDW associated with the IVCT mechanism 2Ag$^{2+}\rightarrow$ Ag$^{1+}+$ Ag$^{3+}$.
Each single branch of the soft mode affects every second layer what is presented in the middle panel of Fig.~\ref{fig2}.
Since both modes make distortions in different layers, they together induce the IVCT in the entire crystal (Fig.~\ref{fig2}, right panel).
The commensurate CDW (C-CDW) instability ($i.e.$ appearing at high-symmetry Z point with commensurate vector $q=(0,0,0.5)$) is accompanied by even more imaginary incommensurate component (I-CDW) that develops along the $\Gamma$-Z direction with minimum around $q=(0,0,0.37)$ and causes the same intra-layer atomic displacements as the C-CDW. The I-CDW mediates the IVCT in all AgF$_2$ layers and simultaneously modulates the stacking of the charge modulated Ag$^{1+}$Ag$^{3+}$Ag$_4$ layers.
Presence of the C-CDW indicates existence of long-range ordering, which means that there is significant inter-planar coupling [52], indicating 3D character of the emerging CDW phase. 
On the other hand, stacking of the CO layers is incommensurate with the underlying crystal lattice in the the I-CDW, which indicates weak inter-planar coupling or even lack of long-range order.

Frequency of the C-CDW mode is highly sensitive to the amplitude of atomic displacements,
showing non-monotonic evolution with increasing amplitude (see Fig.~\ref{fig1}(b)). It first decreases and then increases,
while it remains imaginary only for the intermediate amplitude (u = 0.02~\AA). 
On the other hand, the I-CDW branch along the $\Gamma$-Z direction shows almost no dependence 
on the amplitude and remains imaginary for all calculated values. 
The strong dependence of a phonon frequency on the amplitude of atomic displacement indicates
its strong sensitivity to temperature, which is characteristic of a Kohn anomaly. 
More specifically, it increases with increasing temperature due to the smearing of the
Fermi surface. Therefore, we have explored this dependence using the first order Methfessel-Paxton and
Gaussian smearing functions. Both the C-CDW and the I-CDW modes show high sensitivity to the smearing width
independent of the smearing function, and their frequencies increase with increasing smearing width 
as expected for a Kohn anomaly. The C-CDW mode becomes stabilized (with real frequencies) upon slight increase of
the smearing width from the value of 0.05~eV to 0.1~eV, while the I-CDW mode becomes stabilized at
higher value of 0.2~eV (Fig.~\ref{fig1}(c)). The above results suggest that metallic AgF$_2$ would first transform to an I-CDW phase and subsequently to a C-CDW phase with decreasing temperature: Ag$^{2+}$F$_2$  $\rightarrow$ incommensurate Ag$^{1+/3+}$F$_2$ $\rightarrow$ commensurate Ag$^{1+/3+}$F$_2$. 

\begin{figure}[t!]
\includegraphics[scale=0.21]{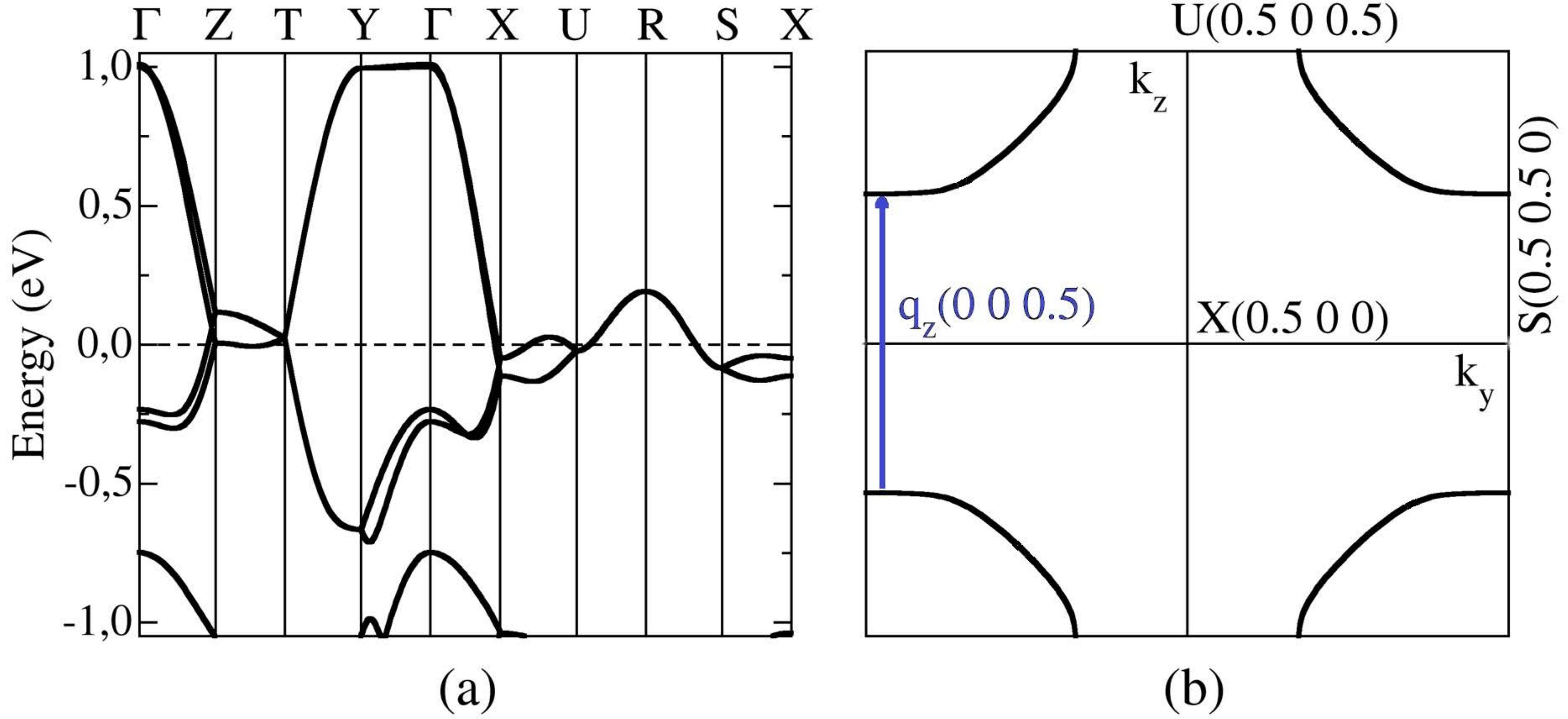}
\caption{Electronic band structure at the vicinity of the Fermi level (a) and
the cross-section of the Fermi surface along the U-X-S line (b) for the metallic AgF$_2$ solution (DFT limit). 
}
\label{Fermi}
\end{figure}

A typical Kohn anomaly is observed at the phonon wave vector $q$, which is related to the Fermi wave vector $k_\text{F}$
by simple relation $q=2k_\text{F}$. 
Such wave vector connects points at the Fermi surface, and the effect is particularly strong if $q$ connects large areas of the Fermi surface parallel to each other. This phenomenon is called nesting, and it plays a role behind the formation mechanism of CDW.
In AgF$_2$, the Fermi surface of the metallic state (in the DFT limit) is composed of 
four bands that are partially filled below the Fermi level (Fig.~\ref{Fermi}(a)). 
Fig.~\ref{Fermi}(b) depicts a cross-section of the Fermi surface of one of these bands with the BZ face defined by the U-X-S plane containing a possible nesting vector $q_\text{Z}=(0, 0, 0.5)$ connecting areas of corner pockets of the Fermi surface. 
In this cross-section, we clearly observe the linear parts of the Fermi surface, which can be connected by the $q_\text{Z}$ wave vector. This is consistent with the Kohn mechanism of the phonon softening at the Z point.

\begin{figure}[t!]
\includegraphics[scale=0.43]{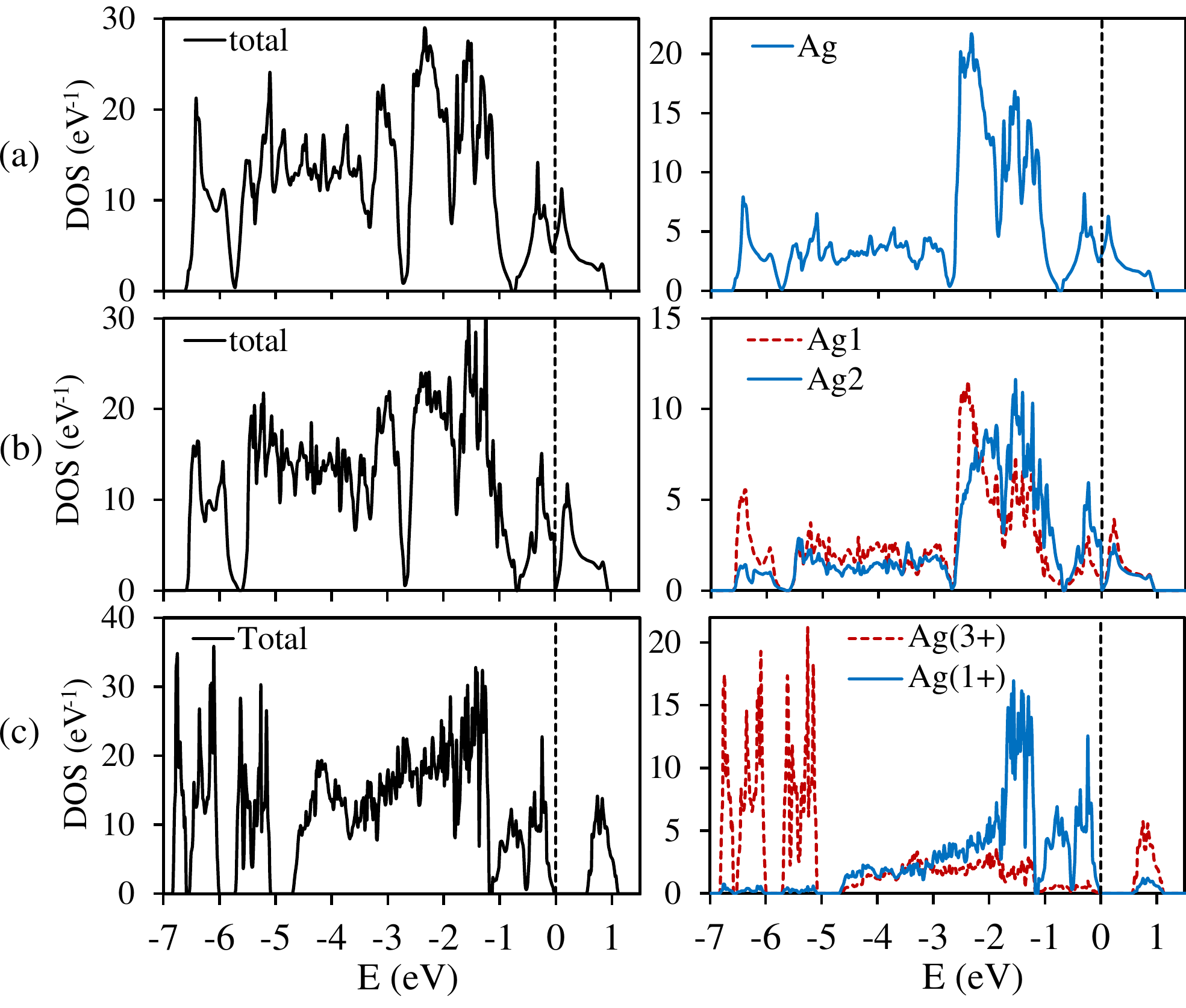}
\caption{Total and atom(Ag)-resolved density of states of undistorted AgF$_2$ in the DFT regime (a), and distorted (optimized) structure along the imaginary Z mode of AgF$_2$ in the DFT (b) and DFT+U regime with $U_\text{Ag}$ = 5 eV (c).}
\label{DOS}
\end{figure}

The charge density redistribution mediated by the soft mode at the Z point is evident from observation of the atom-resolved electronic density of states (DOS) shown in Fig.~\ref{DOS}(b).
It reveals distinct contribution from two neighbouring silver atoms labelled in the DOS plot as Ag1 and Ag2. One can notice decreased population of the lowest-energy valence and conduction states of Ag2 relative to Ag1 that indicates formation of Ag$^{1+}$ ($d^{10}$) and Ag$^{3+}$ (low-spin $d^{8}$), respectively. 
The CDW is accompanied by opening of a point gap at the Fermi level. 
The gap develops in the upper Hubbard Ag($d_{x^2-y^2}$) band that spreads between -1 and 1 eV and it is strongly hybridized with the $p_x$ and $p_y$ orbitals (not shown)~\cite{Grochala2001,Gawraczynski2019}. 
Note that the partial segregation of the $d_{x^2-y^2}$ band to a valence and a conduction band is already present in the undistorted metallic solution near the Fermi level (Fig.~\ref{DOS}(a)). 

It is evident that charge disproportionation is only partial within the DFT picture, since Ag$^{1+}$ cation requires full occupation of all $d$ states, while in the distorted structure both Ag ions have partially depopulated $d$ states. 
The charge disproportionation is further enhanced and a full band gap opens when on-site electron correlations are accounted for. 
The complete charge transfer 2Ag$^{2+}\rightarrow$ Ag$^{1+}+$ Ag$^{3+}$ takes place for realistic value of $U_\text{Ag}$ = 5 eV as witnessed by the atom-resolved DOS in Fig.~\ref{DOS}(c). 
All $d$ states are now fully occupied for Ag2 and the $d_{x^2-y^2}$ band is depopulated for Ag1 as expected for Ag$^{1+}$ and low-spin Ag$^{3+}$, respectively.

We recall that each branch of the doubly degenerate CDW mode at the Z point leads to a local minimum, an intermediate between the AFM and the CO solution, with spin and charge domains distributed to alternating Ag$^{2+}$F$_2$ and Ag$^{1+/3+}$F$_2$ layers (Fig.~\ref{fig2}, middle panel).
In the Ag$^{1+/3+}$F$_2$ layers the band gap opens by charge order while in the Ag$^{2+}$F$_2$ by spin order via the AFM superexchange. The AFM state within the Ag$^{2+}$F$_2$ layers can be obtained only if electron correlations beyond the standard DFT are accounted for~\cite{Kurzydlowski2018,Gawraczynski2019,Tokar2021}. Within DFT+U, it is stabilized for $U\geq$ 1 eV. The band gap in the AFM layers is larger than in the IVCT layers by factor of two irrespective of the  value of $U$. 
The coexistence of spin and charge domains in AgF$_2$ is reminiscent of stripe phases in cuprates characterized by unique distribution of spin and charge densities. In cuprates, however the stripes form within the [CuO$_2$] layers~\cite{Frano2020,Zhang2014}.

It has been suggested that the CDW structure of AgF$_2$ takes on the KBrF$_4$ (Na$^{1+}$Ag$^{3+}$F$_4$) type~\cite{Muller1987,Shen1999,Graudejus2000,Tokar2021}.
This structure requires doubling of the $\alpha$-AgF$_2$ unit cell, which correlates well with 
our theoretical result that reveals the soft mode at the Z point (Fig.~\ref{fig1}(a)).
However, this mode leads to an orthorhombic $Pca2_1$ structure with different ordering of the Ag$^{1+}$ and Ag$^{3+}$ ions and orientation of the square planar [Ag$^{3+}$F$_4$] units than the tetragonal KBrF$_4$ type.
The KBrF$_4$ type structure ensures perfect antiferrodistortive orientation of the units, which minimizes secondary Ag$^{3+}$-F...Ag$^{3+}$ interactions and maximizes splitting between the doubly occupied $d_{z^2}$ and the empty $d_{x^2-y^2}$ orbital. This leads to further broadening of the band gap (compare Fig.~\ref{DOS}(c) with the DOS of the KBrF$_4$ type from reference ~\cite{Tokar2021}) and thus increased stability of the CO state by 27 meV/atom when considering $U_\text{Ag}$ = 5 eV. 
We have found that, when distorting the AgF$_2$ lattice along the Z mode and optimizing it with the hybrid functional, it converges to the KBrF$_4$ type structure.
This result shows that the transition from the spin-ordered ground-state to the charge-ordered ground-state involves strong electron correlations that are not sufficiently captured by the standard DFT and DFT+U functionals.  

The results presented in this study indicate an essential role of lattice dynamics and electron-phonon coupling in CO in AgF$_2$.
The CDW mode at the Z point induces charge disproportionation on Ag ions.
The band splitting in the vicinity of the Fermi level is caused solely by a lattice distortion 
as demonstrated by DFT with no electron correlations included. 
However, the fully charge-ordered state with completely opened bandgap is reached
only when local interactions are taken into consideration within the DFT+U approach. 
Inclusion of exact exchange in form of hybrid DFT functional drives further atomic displacements in the CO state towards the KBrF$_4$-type ground state structure with larger splitting of the the $e_g$ orbitals that provides additional stabilization.
Since the total energy of this phase is higher than the ground state energy of the AFM $\alpha$ phase~\cite{Tokar2021}, 
the $\beta$-AgF$_2$ structure must be treated as metastable.
It explains a fast $\beta\rightarrow\alpha$ transformation observed experimentally~\cite{Shen1999}.

In this contribution we discussed the phonon-driven mechanism of electron localization, which competes with the AFM interactions
and promotes charge disproportionation in AgF$_2$.
We revealed the soft mode related to the Kohn anomaly enhanced by the Fermi surface nesting.
Uncovering the CDW mechanism underneath the AFM spin ordering (that prevails at realistic values of on-site electron correlations), gives us an important hint that one could potentially tune electronic structure between these two regimes (charge-ordering vs. spin ordering) for example by strain or alloying.

The electronic and crystal structure of the studied systems have been optimized using the Density Functional Theory implemented in the Vienna Ab initio Simulation Package (VASP)~\cite{VASP1}. 
For full relaxation of lattice parameters and atomic positions, the projector augmented-wave metod ~\cite{blochl1994} and the generalized gradient approximation optimized for solids~\cite{perdew2008} were used. The energy cut-off was set to 520 eV.
The strong electron interactions in the Ag($4d$) states have been included within the DFT+U method \cite{lichtenstein1995} with the Coulomb parameter $U=5.0$ eV and the Hund's exchange $J=1$ eV, and using the hybrid HSE06 functional with 25$\%$ of exact exchange~\cite{Krukau2006}.
For plotting the Fermi surfaces, the Wannier interpolation was performed on calculated electronic bands using the Wannier90 software~\cite{Pizzi_2020}. 
The Fermi surfaces and their 2D cross sections were plotted by XCrysden~\cite{KOKALJ1999} and Fermisurfer~\cite{Kawamura2019} programs, respectively.
The phonon dispersion curves were obtained using the direct method~\cite{PHONON1} implemented in the program PHONOPY~\cite{phonopy}.

\subsection*{Acknowledgements}

M.D. and K.T. acknowledge the European Regional Development
Fund, Research and Innovation Operational Program (ITMS2014+: 313011 W085), the Slovak Research and Development Agency (APVV-18-0168) and Scientific Grant Agency of the Slovak Republic (VG 1/0223/19). P.P. acknowledges the support by Narodowe Centrum Nauki (NCN, National Science Centre, Poland), Project No. 2017/25/B/ST3/02586. W.G. acknowledges Polish National Science Center (NCN) for Beethoven project (6/23/G/ST5/04320). The research was carried out using Aurel supercomputing infrastructure in CC of Slovak Academy of Sciences (ITMS 26230120002 and 26210120002) funded by ERDF and Interdisciplinary Centre for Mathematical and Computational Modelling (ICM), University of Warsaw (grants no. G62-24, GA76-19, GA67-13).

\bibliography{AgF2}

\end{document}